# Kirchhoff-Law Johnson Noise Meets Web 3.0: A Statistical Physical Method of Random Key Generation for Decentralized Identity Protocols

Christiana Chamon[1,*], Kamalesh Mohanasundar[1], Sarah A. Flanery[2], and Francis K. Quek[3]

[1]Texas A&M University, Department of Computer Science and Engineering, College Station, TX, 77843, USA
[2]Texas A&M University, Department of Electrical and Computer Engineering, College Station, TX, 77843, USA
[3]Texas A&M University, Department of Teaching, Learning, and Culture, College Station, TX, 77843, USA
*cschamon@tamu.edu

## ABSTRACT

This paper presents a statistical physical generation of random keys for a decentralized identity ecosystem that uses Web 3.0 protocols. Web 3.0 is driven by secure keys, typically represented in hexadecimal, that are pseudo-randomly generated by an initialization vector and complex computational algorithms. We demonstrate that the statistical physical Kirchhoff-law-Johnson-noise (KLJN) scheme eliminates the additional computational power by naturally generating truly random binary keys to drive the creation of decentralized identifiers (DIDs) that are appended to an Ethereum blockchain.

Please note: Abbreviations should be introduced at the first mention in the main text – no abbreviations lists. Suggested structure of main text (not enforced) is provided below.

## 1 Introduction[1]

### 1.1 Link to Web 3.0
Web 3.0 currently exists with the purpose of bringing back ownership to its users[1,2,4]. To accomplish this, it takes a decentralized identity approach using blockchain technology[5]. Similar to a linked list, blockchains consist of blocks appended to a preexisting list. Each block contains information about the previous and preceding block. However, no previous blocks can be deleted or altered[6].

Information security in the present day is highly centralized; all data is controlled by singular entities such as the protocols within SSL and TLS, and users have to trust third parties to verify that their data is encrypted[7,8]. For example, social media companies such as Twitter, Facebook, and Instagram are all operated by a centralized power: the admin assigned by each company. This system of storing information is based on Web 2.0. The basis of Web 2.0 is that a central power ensures the safety and accuracy of information online and decides what is distributed to its users. These administrators control what information is distributed to its users and which must be hidden or deleted from the platform. For these reasons, this has the potential to be a gateway to corruption, as the basis of the population's access to information are influenced by personal biases[9].

Web 3.0, due to the nature of its data structure, eliminates this problem entirely by removing the centralized power and using other methods of verification to determine the accuracy of information on the internet. The obstacle encountered when trying to incorporate Web 3.0 into education is the centralized nature of learning. In modern society, the validity of an academic credential is established by the institution itself rather than defining individual skills learned. In Web 2.0, the user does not have any ownership of the information that they post online. However, Web 3.0 resolves this issue. In the present paper, we demonstrate that the decentralized elements of Web 3.0 can be utilized in order to ensure that the accomplishments in education remain assigned to the user.

### 1.2 Decentralized Identity
Decentralized Identity is a form of authentication standardized by Web 3.0 that does not require a major entity to validate the user[10]. In the current state of the internet, the user's identity is tied to the centralized source. For example, most academic applications involve having to utilize an educational account that is created by an external provider. If these external providers were to become deactivated, these personal accounts would be displaced. Decentralized identity attempts to resolve this by using an Ethereum blockchain to append blocks tied to each user's specific decentralized identifier (DID). Centralized figures are unable to regulate DID users, and this provides ownership of the user's access and possession of online data. Increased ownership of information, similar to the basis of Web 3.0, has rendered DID a normalized method of self-identity. For example, the European Union (EU) has been trying to adopt DID in order to replace other forms of government ID such as passports and driver's licenses[11]. Citizens can not only have all their necessary identification in a singular place, but also as a method to decrease potential government corruption.

---
[1]Copied and adapted from[1–3]

DID is generated using asymmetric cryptography, consisting of a private key pair and a hash that are computationally generated in Ethereum. The DID document containing the identity of the user is then verified by either an outside entity or the user's self. In the latter case, biometric login such as fingerprinting and face ID is often used, as it cannot be easily replicated by outside persons[12]. Once verified, the DID document, such as the one displayed in Figure 1, is appended to a distributed ledger, often a blockchain. However, blockchain architecture is not required to perform this. Decentralized Identifiers allow users to take back ownership of their online credentials, as there is no centralized figure managing the ecosystem.

```
DID                                                                                    Alias
did:we  ┌did:ethr:0x025f2c19148f0afb66ef9f3c7cc65464ba2d4e9339784d1a6a7239c059e1f75014─┐
did:et  {
did:we      "did": "did:ethr:0x025f2c19148f0afb66ef9f3c7cc65464ba2d4e9339784d1a6a7239c059e1f75014",
did:et      "provider": "did:ethr",
did:et      "alias": "Kamalesh Mohanasundar",
did:et      "controllerKeyId": "045f2c19148f0afb66ef9f3c7cc65464ba2d4e9339784d1a6a7239c059e1f750149c8a5997fe2f713eb8d31db93b5b5e8716332
did:et  fb17a9333643d3fd3f5748cbfc0",
did:et      "keys": [
did:et          {
did:et              "kid": "045f2c19148f0afb66ef9f3c7cc65464ba2d4e9339784d1a6a7239c059e1f750149c8a5997fe2f713eb8d31db93b5b5e8716332fb17a933
did:et  3643d3fd3f5748cbfc0",
did:et              "kms": "local",
did:et              "type": "Secp256k1",
did:et              "publicKeyHex": "045f2c19148f0afb66ef9f3c7cc65464ba2d4e9339784d1a6a7239c059e1f750149c8a5997fe2f713eb8d31db93b5b5e871633
did:et  2fb17a9333643d3fd3f5748cbfc0",
                   "meta": {
                       "algorithms": [
                           "ES256K",
                           "ES256K-R",
                           "eth_signTransaction",
```

**Figure 1.** Typical DID document for alias "Kamalesh Mohanasundar" consisting of the domain of the ledger ("did:ethr"), the user's specifically-assigned DID (given by "did"), and respective public key (given by "publicKeyHex")[2].

The DID keys are generated using the base encryption methods for Ethereum blockchains, i.e. a combination of public and private key encryption[13]. The public key can be generated using a public key hex and ES256 algorithms[14]. The private key is generated based on elliptic curve cryptography which is commonly used in Bitcoin[14,15]. However, the vulnerability with the pseudo-random generation is that if either of the keys are found, anyone can access the user. Additionally, these pseudo-random methods involve multiple computational steps, thus the amount of computational power increases accordingly. We believe that computers are not necessary to generate these keys, i.e. the keys can be generated naturally. In the present paper, we explore how the statistical physical phenomena of the Kirchhoff-law-Johnson-noise (KLJN) secure key exchange scheme[3, 16–40, 40–80] can generate keys to support the Semantic Web protocols that drive decentralized identifiers.

### 1.3 The KLJN Scheme
The KLJN scheme is a statistical physical scheme based on the second law of thermodynamics. Figure 2 illustrates the core of the KLJN scheme. Communicating parties Alice and Bob are connected by a publicly accessible wire, which has voltage and current, denoted by $U_w(t)$ and $I_w(t)$, respectively. Each communicating party has identical pairs of resistors $R_H$ and $R_L$ ($R_H > R_L$) and their respective artificial noise generators $U_{H,A}(t)$, $U_{L,A}(t)$, $U_{H,B}(t)$, and $U_{L,B}(t)$ (the former two belonging to Alice, the latter two belonging to Bob) that emulate thermal noise at a publicly-agreed-upon $T_{\text{eff}}$ (typically $> 10^{15}$ K).

To generate a single bit, Alice and Bob begin the bit exchange period (BEP) by randomly selecting one of their resistors to connect to the wire, observing the instantaneous noise voltages over the wire, and taking the mean-square value of those noise voltages to assess the bit status. The mean-square voltage is given by the Johnson Formula,

$$U_w^2 = 4kT_{\text{eff}}R_p \Delta f_B, \tag{1}$$

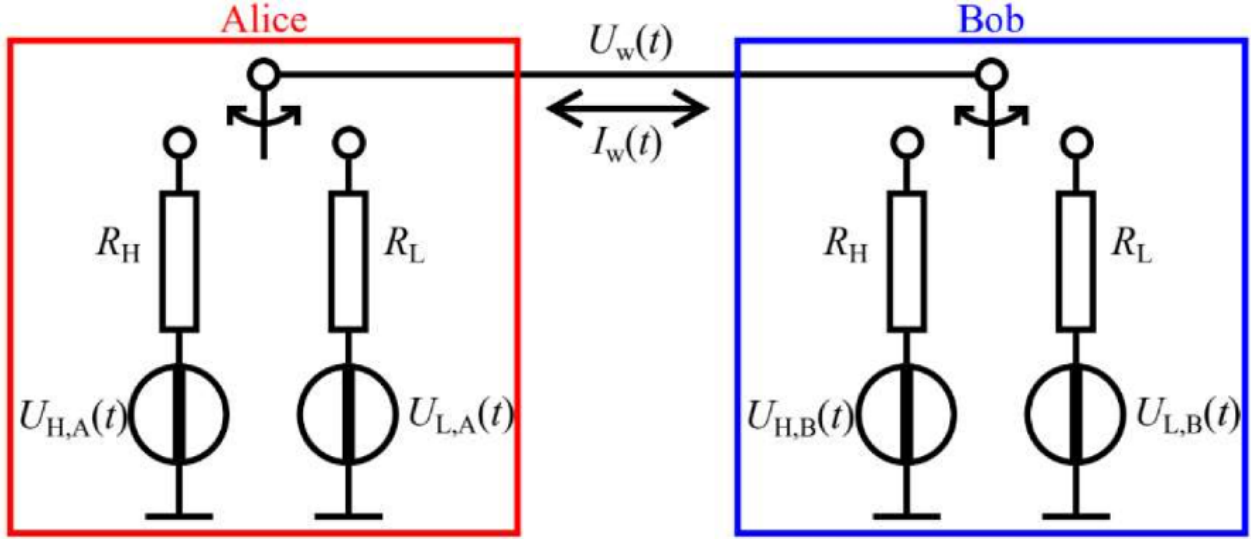

**Figure 2.** The core of the KLJN scheme. The two communicating parties, Alice and Bob, are connected via a wire. The wire voltage and current are denoted as U$U_w(t)$ and $I_w(t)$, respectively. The parties have iidentical pairs of resistors $R_H$ and $R_L$ ($R_H > R_L$) that are randomly selected and connected to the wire at the beginning of the bit exchange period. The statistically independent thermal noise voltages $U_{H,A}(t)$, $U_{L,A}(t)$, $U_{H,B}(t)$, and $U_{L,B}(t)$ represent the noise voltages of the resistors $R_H$ and $R_L$ of Alice and Bob, respectively[21,24,26,28].

where $k$ represents the Boltzmann constant ($1.38 \times 10^{-23}$ J/K), $T_{eff}$ represents the publicly-agreed-upon effective temperature, $R_p$ represents the parallel combination of Alice's and Bob's chosen resistors, given by

$$R_p = \frac{R_A R_B}{R_A + R_B} \quad (2)$$

where $R_A$ represents Alice's resistor choice and $R_B$ represents Bob's resistor choice, and $Delta f_B$ represents the noise bandwidth of the generators.

Four possible resistance combinations can be formed by Alice and Bob: HH, LL, LH, and HL. Using the Johnson Formula, these correspond to three mean-square voltages, as shown in Figure 3. The HH and LL cases are insecure situations because they render a distinct mean-square voltage; these bits are discarded by Alice and Bob. The LH and HL cases, on the other hand, are secure situations because they render the exact same mean-square voltage, thus an adversary cannot differentiate between the two situations; however, Alice and Bob know which resistor they have chosen.

We typically represent the LH case as the bit value "0" and the HL as the bit value "1"[22,75]. Over the course of several bit exchange periods, Alice and Bob generate a secure binary key, or random string of bits (see Section 1.2). All binary keys can be converted into a hexadecimal representation, thus we believe that a system that utilizes hex keys can be driven by converting the base of the original binary keys. In the present paper, we explore a proof of concept involving the utilization of the statistical physical phenomena of the KLJN scheme to provide secure keys to drive a DID credentialing system that we created[2].

The rest of this paper is as follows. Section 2 describes the methodology, Section 3 demonstrates the KLJN-powered identifier, and Section 4 concludes this paper.

## 2 Methodology

Alice and Bob randomly select one of their resistors, $R_H$ or $R_L$, to connect to the wire (see Section 1.3). If both communicating parties chose $R_H$ or $R_L$ (i.e. the HH or the LL case), we discard the bit; however, in the secure bit situations where Alice chose $R_H$ and Bob chose $R_L$ (the HL case), or where Alice chose $R_L$ and Bob chose $R_H$ (the LH case), we keep the bit.

Alice and Bob undergo as many bit exchange periods as needed to produce a secure binary key of a fixed length $L$. We employ base conversion to convert the binary key to its respective hexadecimal representation, which contains $\frac{1}{4}L$ characters. We then use the resulting hex key to generate a DID for a single user and append the DID to an Ethereum blockchain.

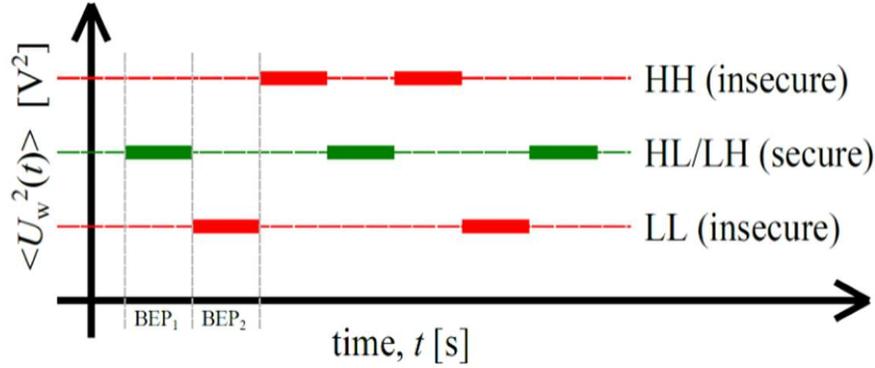

**Figure 3.** The three mean-square voltage levels. The HH and LL cases represent insecure situations because they form distinct mean-square voltages. The HL and LH cases represent secure bit exchange because Eve cannot distinguish between the corresponding two resistance situations (HL and LH). On the other hand, Alice and Bob can determine the resistance at the other end because they know their own connected resistance value[21, 24, 26, 28].

## 3 Demonstration

### 3.1 Statistical Key Generation

#### 3.1.1 Johnson Noise Emulation[2]

First, we generated Gaussian band-limited white noise (GBLWN). Precautions were used to avoid aliasing errors, improve Gaussianity, and reduce bias:

1. At first, using the MATLAB randn() function, 224 or 16,777,216 Gaussian random numbers were generated.

2. This process was repeated 10 times to generate 10 independent raw noise series, and then an ensemble average was taken out of those 10 series to create one single noise time function with improved Gaussianity and decreased short-term bias.

3. Then this time series was converted to the frequency domain by a Fast Fourier Transformation (FFT). To get rid of any aliasing error, we opened the real and imaginary portions of the FFT spectrum and doubled their frequency bandwidths by zero padding to represent Nyquist sampling.

4. Finally, we performed an inverse FFT (IFFT) of the zero-padded spectrum to get the time function of the anti-aliased noise. The real portion of the IFFT result is the band-limited, antialiased noise with improved Gaussianity and decreased bias.

The probability plot of the generated noise is shown in Figure 4, showing that the noise is Gaussian. Figure 5 demonstrates that the noise has a band-limited, white power density spectrum and that it is anti-aliased.

From the Nyquist Sampling Theorem,

$$\tau = \frac{1}{2\Delta f_B} \tag{3}$$

where $\tau$ represents the time step, an $\Delta f_B$ of 500 Hz renders a time step of $10^{-3}$ seconds.

The final step was to scale this normalized Gaussian noise to the required level of Johnson noise for the given resistance, bandwidth, and temperature values ($R_H$, $R_L$, $\Delta f_B$, and $T_{\text{eff}}$, respectively). We chose $R_H = 100$ kΩ, $R_L = 10$ kΩ, $\Delta f_B = 500$ Hz and $T_{\text{eff}} = 10^{18}$ K).

A realization of Alice's and Bob's noise voltages over 100 milliseconds is displayed in Figure 6. $U_{H,A}(t)$ is the noise voltage of Alice's $R_H$, $U_{L,A}(t)$ is the noise voltage of Alice's $R_L(t)$, $U_{H,B}(t)$ is the noise voltage of Bob's $R_H(t)$, and $U_{L,B}(t)$ is the noise voltage of Bob's $R_L(t)$ (see Figure 2). Each time step is one millisecond[21, 24, 26, 28].

---

[2]Copied and adapted from[3]

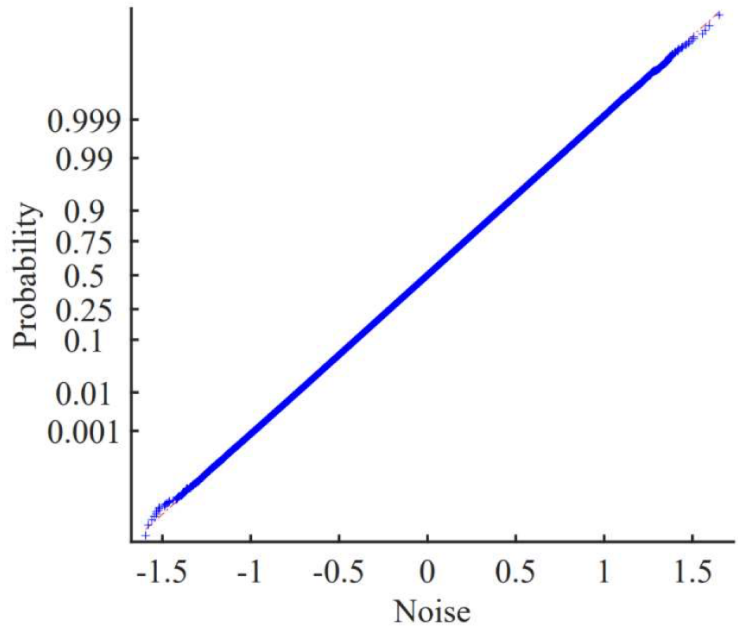

**Figure 4.** Normal-probability plot of the noise[21,24,26,28]. A straight line indicates a pure Gaussian distribution.

### 3.1.2 Bit Exchange

To simulate Alice and Bob randomly selecting one of their resistors (see Section 1.3), we used the MATLAB randi() function for each communicating party to randomly generate 0s and 1s. If the output was 0, the chosen resistor for the respective communicating party was $R_L$, and if the output was 1, the chosen resistor for the respective communicating party was $R_H$. In the insecure bit situations where Alice and Bob both produced a 0 or 1 (i.e. the LL or the HH case), the bit was discarded; however, in the secure bit situations where Alice produced a 0 and Bob produced a 1 (the LH case), or where Alice produced a 1 and Bob produced a 0 (the HL case), the bit was kept.

We simulated 256 bit exchanges to produce a key of $L = 256$ bits (see Section 2). The corresponding hex representation contains $\frac{L}{4} = 64$ characters.

## 3.2 DID Setup[3]

We used Veramo to issue DIDs and credentials because of its open-source framework that facilitates creating and managing digital identities[19,20]. By taking advantage of Veramo's framework and modular design, we are able to incorporate the decentralized identity management and credential verification functionalities to successfully generate DIDs (see Section 1.2).

We also linked Veramo to OpenSSL for the purpose of providing an extra layer of security in comparison to using Veramo alone. Using the OpenSSL GitHub repository, we were able to establish secure connections between web servers[16]. External security in a decentralized environment ensures that operations between systems are not maliciously infiltrated.

Integration into a demo web app involved installing Node dependencies such as "yarn" and creating the agent within a nodeJS app. Because the plugins Veramo uses are all public and standardized, any verifier can resolve DIDs.

## 3.3 Key Integration

In order for Veramo to be able to properly read the KLJN-generated key (see Section 3.1.2), we created the "binaryToHex" function to regroup every four binary digits into a single hexadecimal character. Then, we input the resulting hex key into Veramo's Key Management System, and we created an identifier by entering the command "yarn ts-node –esm ./src/create-identifier.ts"[2]. An example of a resulting DID is displayed in Figure 4. The identifier has a unique DID string that is associated with a key pair. The identifier also contains Ethereum-specific signing functions such as "ES256K", "ES256K-R" , and "eth_signTransaction that define how the keys can be used for transactions and interactions on the Ethereum blockchain.

---

[3]Copied and adapted from[2]

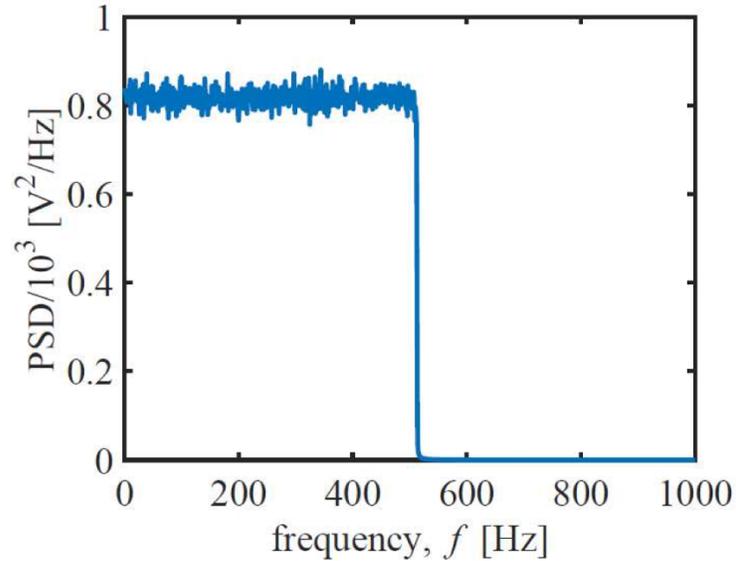

**Figure 5.** Power spectral density of the noise[21, 24, 26, 28]. The bandwidth of the noise is 500 Hz, see (Equation (1)).

## 4 Conclusion

Decentralized identity ecosystems depend on secure keys to function. Such keys are represented in hexadecimal, and they rely on computational algorithms for production. We showed that such keys can be generated in the absence of a computer, i.e. with statistical physical means. Since the keys are truly random, they are immune to vulnerabilities involving computational knowledge. We can then employ a base conversion algorithm and use the output to create a decentralized identifier that was appended to an ethereum blockchain.

It is important to note that this work was carried out on a local client-server network for a single user. Future work would involve integration on a cloud server and mass scalability of the KLJN scheme to make it accessible for multiple users.

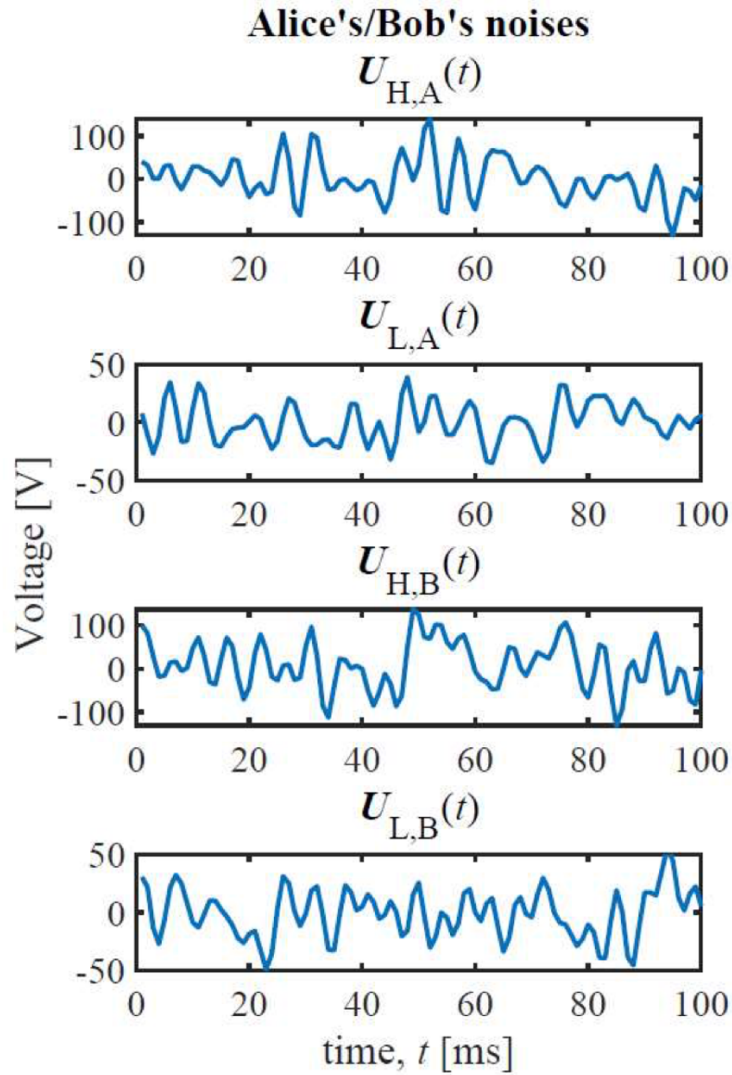

**Figure 6.** A realization of $U_{H,A}(t)$, $U_{L,A}(t)$, $U_{H,B}(t)$, and $U_{H,B}(t)$ (see Figure 2) for Alice and Bob, displayed over 100 milliseconds[21,24,26,28].

```
kamalesh_mohanasundar@Kamaleshs-MBP veramo-agent % yarn ts-node --esm ./src/create-identifier.ts
(node:48674) ExperimentalWarning: Import assertions are not a stable feature of the JavaScript language. Avoi
d relying on their current behavior and syntax as those might change in a future version of Node.js.
(Use `node --trace-warnings ...` to show where the warning was created)
(node:48674) ExperimentalWarning: Importing JSON modules is an experimental feature and might change at any t
ime
Binary Key: 1111001110100100000101111000001111011100111110111101100011001110010011001001000100100001011111
00101011001010100011001001010110110010111000110001111111110111001000011111000011001011101000001111110111111010
0111100011100011101110110011010101000101010001101
Hex Key: f3a41783dcfdec679324482f9595192b65c63fdc87c65d07efd3c78eecd5150d
New identifier created
{
  "did": "did:ethr:goerli:0x0377914794f5cc4345c54a8b2374445593d1cc16c912e6ed0302217ee7432fbbb3",
  "controllerKeyId": "0477914794f5cc4345c54a8b2374445593d1cc16c912e6ed0302217ee7432fbbb32b32ed8d951970a9e5b44
7af9855547408786d8aa65710752c731d0e1dbe4665",
  "keys": [
    {
      "type": "Secp256k1",
      "kid": "0477914794f5cc4345c54a8b2374445593d1cc16c912e6ed0302217ee7432fbbb32b32ed8d951970a9e5b447af98555
47408786d8aa65710752c731d0e1dbe4665",
      "publicKeyHex": "0477914794f5cc4345c54a8b2374445593d1cc16c912e6ed0302217ee7432fbbb32b32ed8d951970a9e5b4
47af9855547408786d8aa65710752c731d0e1dbe4665",
      "meta": {
        "algorithms": [
          "ES256K",
          "ES256K-R",
          "eth_signTransaction",
          "eth_signTypedData",
          "eth_signMessage",
          "eth_rawSign"
        ]
      },
      "kms": "local"
    }
  ],
  "services": [],
  "provider": "did:ethr:goerli",
  "alias": "Binary test 6"
}
```

**Figure 7.** Example of a generated identifier consisting of the binary key as its base, the respective hex representation as its operator, the essential metadata (i.e. the Key Identifier, Public Key, Type of Cryptographic Key, and The Key Management System), and Ethereum-specific signing functions to enable communication with the Ethereum blockchain.